# Stability, structural, elastic and electronic properties of polymorphs of the superconducting disilicide YIr$_2$Si$_2$

I.R. Shein

*Institute of Solid State Chemistry, Ural Branch of the Russian Academy of Sciences, 620990, Ekaterinburg, Russia*

**Abstract**

Using the *ab initio* approaches, the comparative stability, structural, elastic, and electronic properties of three polymorphs of the superconducting disilicide YIr$_2$Si$_2$, which differ in the atomic configurations of [Ir$_2$Si$_2$] (or [Si$_2$Ir$_2$]) blocks, were examined. For these YIr$_2$Si$_2$ polymorphs, the optimized structural data, elastic parameters, electronic bands, total and partial densities of states, Fermi surface topology, and chemical bonding have been obtained and analyzed. Our studies showed that although ThCr$_2$Si$_2$- and CaBe$_2$Ge$_2$-type polymorphs are mechanically stable and relatively hard materials with low compressibility, they will behave as ductile systems. Among them, ThCr$_2$Si$_2$-type polymorph will show enhanced elastic anisotropy. In the vicinity of the Fermi energy, the topology of the electronic bands and the Fermi surface for various polymorphs are quite different. Besides, the CaBe$_2$Ge$_2$-type polymorph is expected to be anisotropic, *i.e.* happening mainly in the [Si$_2$Ir$_2$] blocks. The inter-atomic bonding for YIr$_2$Si$_2$ polymorph phases can be described as an anisotropic mixture of covalent, metallic, and ionic contributions, where inside the [Ir$_2$Si$_2$] (or [Si$_2$Ir$_2$]) blocks, Ir-Si and Ir-Ir bonds take place, whereas between the adjacent [Ir$_2$Si$_2$] (or [Si$_2$Ir$_2$]) blocks and Y atomic sheets, Si-Si and Ir-Y, Si-Ir and Si-Y, or mainly Ir-Ir bonds emerge for various polymorphs.

\* *E-mail address:* shein@ihim.uran.ru

# 1. Introduction

The discovery [1] of superconductivity in close proximity to magnetism in very intriguing iron-based layered materials (with maximal $T_C$ to 55K) attracted tremendous attention to several related groups of compounds, reviews [2-7]. Among them, so-called 122 phases belong to one of the most interesting and well studied groups of such materials. In turn, these 122 iron-based superconductors (SCs) include so-called iron-pnictide phases *AE*Fe$_2$*Pn*$_2$ (where *AE* are alkali-earth metals with $T_C$ to 38K) [7-12] and the newest family of iron-chalcogenide SCs *A*Fe$_2$Se$_2$ (*A* = alkali metals or Tl; with $T_C$ to 35K), see [13-15].

All these 122-like phases adopt a tetragonal layered ThCr$_2$Si$_2$-type structure with space group *I*4/*mmm* (# 139), which includes quasi-two-dimensional [Fe$_2$*Pn*$_2$] (or [Fe$_2$Se$_2$]) blocks, which are separated by *AE* (or *A*) atomic sheets. The atomic substitutions inside [Fe$_2$*Pn*$_2$] (or [Fe$_2$Se$_2$]) blocks exert a profound influence on the properties of these 122 phases [2-12]. Recently, a rich set of more complex *A*Fe$_{2-x}$*M*$_x$*Pn*$_2$ phases doped with heavy 4*d*, 5*d* metals (*M* = Pd, Ru, Rh, or Ir) was synthesized and their properties were investigated, see [2-12,16-21]. Moreover, the related phase: iron-free SC SrPt$_2$As$_2$ ($T_C$ ~ 5.2K), which may be viewed as the fully Pt-substituted *AE*Fe$_2$*Pn*$_2$, was discovered recently [22,23]. The most remarkable feature of this phase is a very intriguing structural situation: as distinct from the aforementioned ThCr$_2$Si$_2$-like 122 phases, SrPt$_2$As$_2$ adopts a CaBe$_2$Ge$_2$-type structure (space group *P*4/*nmm*, # 129) with an "inverse" type of quasi-two-dimensional [As$_2$Pt$_2$] blocks.

Let us note that besides SrPt$_2$As$_2$, a rich set of 122-like phases with tetragonal CaBe$_2$Ge$_2$-type structures was prepared (SrNi$_2$(Sb,Bi)$_2$ [24]; *RM*$_2$*X*$_2$ compounds, *R* = Th, Np, Pu; *M* = some of 3*d*, 4*d* or 5*d* metals; and *X* = Si, Ge [25]; (Th,U)(Co,Ni,Cu)$_2$Sn$_2$ [26], EuZn$_2$Sn$_2$ [27], *etc*.). Some of them were found to wxhibit low-temperature superconductivity, see [28,29].

Besides, some of 122-like compounds may adopt either ThCr$_2$Si$_2$- or CaBe$_2$Ge$_2$-type structures – depending on synthesis conditions. Such systems seem to be the ideal materials for the study of the interplay of structural peculiarities *versus* physical properties of 122-like phases. These systems include the ternary disilicide YIr$_2$Si$_2$, for which two superconducting polymorphic forms exist: a low-temperature modification (LT) with a ThCr$_2$Si$_2$-type structure ($T_C$ < 2K) and a high-temperature modification (HT) with a CaBe$_2$Ge$_2$-type structure ($T_C$ ~ 2.8K) [28,30,31].

In the present work, by means of the first-principles calculations, we analyze in details the structural, electronic, and elastic properties of the two aforementioned LT and HT polymorphs of YIr$_2$Si$_2$ as a function of their atomic structure and inter-atomic bonding. In addition, a similar analysis for a hypothetical "intermediate" YIr$_2$Si$_2$ polymorph was performed.

## 2. Structural models and computational aspects.

The synthesized [30,31] LT polymorph of YIr$_2$Si$_2$ (termed further as YIS-1) crystallizes in a tetragonal ThCr$_2$Si$_2$-type structure, where the atomic positions are Y: 2$a$ (0, 0, 0), Ir: 4$d$ (0, ½, ¼), and Si: 4$e$ (0, 0, $z_{Si}$). The structure of this phase can be schematically described as a sequence of Y sheets and [Ir$_2$Si$_2$] blocks consisting of {IrSi$_4$} tetrahedrons: ..[Ir$_2$Si$_2$]/Y/[Ir$_2$Si$_2$]/Y/[Ir$_2$Si$_2$]/Y… as shown in Fig. 1. In turn, CaBe$_2$Ge$_2$-type HT polymorph (termed as YIS-2) may be described as a sequence of Y sheets and [Ir$_2$Si$_2$] and [Si$_2$Ir$_2$] blocks consisting of {IrSi$_4$} and {SiIr$_4$} tetrahedrons, respectively: …[Ir$_2$Si$_2$]/Y/[Si$_2$Ir$_2$]/Y/[Ir$_2$Si$_2$]/Y/[Si$_2$Y]…, see Fig. 1. Here, the atomic positions are Y 2$c$ (¼, ¼, $z_Y$), Ir$^1$ 2$a$ (¾, ¼, 0), Ir$^2$: 2$c$ (¾, ¾, $z_{Ir}$), Si$^1$: 2$b$ (¾, ¼, ½), and Si$^2$: 2$c$ (¾, ¾, $z_{Si}$), where $z_{Y,Ir,Si}$ are the so-called internal coordinates. Finally, we also used the ThCr$_2$Si$_2$-type structure as an additional "intermediate" hypothetical polymorph (abbreviated as YIS-3), which has however an inverse distribution of Ir and Si over the atomic sites (in blocks) as compared with YIS-1, *i.e.* the atomic positions here are: Y: 2$a$ (0, 0, 0), Si: 4$d$ (0, ½, ¼), and Ir: 4$e$ (0, 0, $z_{Ir}$). The stacking sequence for YIS-3 is … Y/[Si$_2$Ir$_2$]/Y/[Si$_2$Ir$_2$]/Y…, see Fig. 1. As a result, three basic tetragonal types of YIr$_2$Si$_2$ polymorphs have been examined, which enable us to clarify the role of local atomic arrangement inside [Ir-Si] blocks, *i.e.* {IrSi$_4$} *versus* {SiIr$_4$}, and the role of the main types of stacking of these blocks, when the external atoms from neighboring blocks can form various inter-blocks bonds, namely, Si-Ir, Si-Si, and Ir-Ir for YIS-1, YIS-2, and YIS-3, respectively.

Our calculations were carried out by means of the full-potential method with mixed basis APW+lo (LAPW) implemented in the WIEN2k suite of programs [32] in a scalar-relativistic approximation. The generalized gradient correction (GGA) to exchange-correlation potential in the PBE form [33] was used. The plane-wave expansion was taken to $R_{MT} \times K_{MAX}$ equal to 8, and the $k$ sampling with 12×12×12 $k$-points in the Brillouin zone was used. The MT sphere radii were chosen to be 2.25 a.u. for Ir, 2.0 a.u. for Si, and 2.5 a.u. for Y. The calculations were performed with full-lattice optimization including internal coordinates. The self-consistent calculations were considered to be converged when the difference in the total energy of the crystal did not exceed 0.1 mRy and the difference in the total electronic charge did not exceed 0.001 $e$ as calculated at consecutive steps.

The hybridization effects were analyzed using the densities of states (DOSs), which were obtained by the modified tetrahedron method [34]; some peculiarities of intra-atomic bonding picture were also visualized by means of charge density maps

Furthermore, for the calculations of the elastic parameters for YIr$_2$Si$_2$ polymorphs we employed the Vienna *ab initio* simulation package (VASP) in projector augmented waves (PAW) formalism [35,36]. Exchange and correlation were described by a nonlocal correction for LDA in the form of GGA [33]. The kinetic energy cutoff of 500 eV and k-mesh of 14×14×7 were used. The geometry optimization was performed with the force cutoff of 1 meV/Å.

These two DFT-based codes are complementary and allowed us to perform a complete investigation of the declared properties of the above systems.

**3. Results and discussion**

*3.1. Structural properties and stability*

As the first step, the total energies ($E_{tot}$) *versus* the cell volume were calculated to determine the equilibrium structural parameters for YIr$_2$Si$_2$ polymorphs; see Table 1. The obtained data are in reasonable agreement with the available experiments [28]. Some divergences are related to the well-known overestimation of the lattice parameters within LDA-GGA based calculation methods. The calculated parameters allow us to make the following conclusions.

The total-energy differences ($\Delta E$) between the examined YIr$_2$Si$_2$ polymorphs reveal that both within the FLAPW and VASP approaches the most stable is LT YIS-1 polymorph with a ThCr$_2$Si$_2$-type structure, whereas the HT phase (CaBe$_2$Ge$_2$-like, YIS-2) is by about 0.1 eV/f. u. less stable. Finally, the hypothetical "intermediate" polymorph YIS-3 becomes the most unstable system.

For the ThCr$_2$Si$_2$-type YIS-1 polymorph, the external planes of [Ir$_2$Si$_2$] blocks are formed by Si atoms, and the nearest Si-Si distances between neighboring blocks (~2.4 Å) are much smaller than the corresponding inter-atomic distances inside the blocks (~3.9 Å). On the contrary, for the CaBe$_2$Ge$_2$-type polymorph (YIS-2), the Ir-Si bond lengths (~2.43 Å) between neighboring [Ir$_2$Si$_2$]/[Ir$_2$Si$_2$] blocks are comparable with those for the Ir-Si lengths (about 2.38 Å) inside these blocks. Thus, this polymorph may be viewed as quite an isotropic phase with a three-dimensional (3D) system of strong Ir-Si bonds. In turn, for the hypothetical polymorph YIS-3, the nearest Ir-Ir distances (~3.8 Å) inside the blocks are much larger than the Ir-Ir inter-atomic distances (~2.6 Å) between neighboring blocks. These simple crystallographic reasons allow us to expect that in contrast to YIS-2 with a 3D-like system of Ir-Si bonds, the bonding for YIS-1 and YIS-3 should be very anisotropic: simultaneously with Ir-Si bonds inside the blocks, direct Si-Si (for YIS-1) or Ir-Ir bonds (for YIS-3) will be formed between the corresponding blocks, see also below.

*3.2. Elastic properties*

Let us discuss the elastic parameters for the YIr$_2$Si$_2$ polymorphs as obtained within VASP calculations. The standard "volume-conserving" technique was used in the calculation of stress tensors on strains applied to the equilibrium structure to obtain the elastic constants $C_{ij}$. In this way the values of six independent elastic constants for tetragonal crystals ($C_{11}$, $C_{12}$, $C_{13}$, $C_{33}$, $C_{44}$, and $C_{66}$) were estimated, Table 2.

First of all, $C_{ij}$ constants for the YIS-1 and YIS-2 polymorphs are positive and satisfy the generalized criteria [37] for mechanically stable tetragonal materials: $C_{11} > 0$, $C_{33} > 0$, $C_{44} > 0$, $C_{66} > 0$, $(C_{11}-C_{12}) > 0$, $(C_{11}+ C_{33} - 2C_{13}) > 0$, and $[2(C_{11} + C_{12}) + C_{33} + 4C_{13}] > 0$. On the contrary, for YIS-3 the value of $C_{66}$ is negative. Thus, this polymorph (which is also energetically most unstable, Table I) belongs to mechanically unstable systems, and further in this section it will not be discussed.

Further, the calculated elastic constants $C_{ij}$ allowed us to obtain the bulk ($B$) and shear ($G$) moduli. Usually, for such calculations two main approximations are used, namely the Voigt (V) and Reuss (R) schemes, see for example Ref. [38]. We evaluated also the corresponding parameters for polycrystalline $SrPt_2As_2$ species, *i.e.* for materials in the form of aggregated mixtures of microcrystallites with random orientation. For this purpose we utilized the Voigt-Reuss-Hill (VRH) approximation, see [39]. In this approach, the actual effective moduli ($B_{VRH}$ and $G_{VRH}$) for polycrystals are approximated by the arithmetic mean of the two above mentioned – (Voigt and Reuss) limits and further allowed us to obtain the Young's moduli $Y$ and the Poisson's ratio $v$ as: $Y_{VRH} = 9\ B_{VRH}/\{1 + (3B_{VRH}/G_{VRH})\}$, and $v = (3B_{VRH} - 2G_{VRH})/2(3B_{VRH} + G_{VRH})$.

The above elastic parameters are presented in Table 3. We see that the bulk and shear moduli of the $YIr_2Si_2$ polymorphs increase in the sequence: YIS-1 < YIS-2. As these moduli represent the resistance of crystal to external forces, this reflects probably a more isotropic (3D-like) inter-atomic bonding situation for YIS-2. Besides, it was found that $B > G$; this implies that the parameter limiting the mechanical stability of these materials is the shear modulus $G$.

The obtained bulk moduli for both polymorphs are close to 200 GPa. Thus, $YIr_2Si_2$ should be classified (unlike related 122-like Fe-based phases [40-43]) as a relatively hard material [44] with low compressibility ($\beta \sim 0.0056$ GPa$^{-1}$). In addition, the Young's modulus of materials is defined as a ratio of linear stress and linear strain, which is indicative of their stiffness. The Young's modulus of $YIr_2Si_2$ was found to be $Y \sim 175$ GPa; thus, this material will show a rather high stiffness.

One of the most widely used malleability measures of materials is the Pugh's criterion ($G/B$ ratio) [45]. As is known, if $G/B < 0.5$, a material behaves in a ductile manner, and *vice versa*, if $G/B > 0.5$, a material demonstrates brittleness. In our case, according to this indicator, epy $YIr_2Si_2$ polymorphs will behave as ductile materials.

Elastic anisotropy of crystals reflects a different character of bonding in different directions and has an important implication since it correlates with the possibility to induce microcracks in materials, see for example [46]. We have estimated the elastic anisotropy for the examined materials using the so-called universal anisotropy index [47] defined as: $A^U = 5G_V/G_R + B_V/B_R - 6$. For isotropic crystals $A^U = 0$; the deviations of $A^U$ from zero define the extent of crystal anisotropy. In our case, the minimal anisotropy is exhibited by the 3D-like YIS-2, while the YIS-1 polymorph demonstrates a deviation from $A^U = 0$ testifying to enhanced elastic anisotropy of this system.

*3.3. Electronic band structure and Fermi surface*

The calculated band structure and electronic densities of states (DOSs) for the considered $YIr_2Si_2$ polymorphs are shown in Figs. 2 and 3, respectively. For all the polymorphs, their electronic spectra show some common features, namely (i) the Si 3$s$ states occur between -12 eV and -7 eV with respect to the Fermi energy (E$_F$= 0 eV); (ii) most of the bands between -6.5 eV and E$_F$ are mainly of a mixed Ir 5$d$ + Si 3$p$ character; and (iii) the contributions from the valence states of Y to the occupied bands are quite small. However, in the vicinity of the Fermi energy the topology of the electronic bands for various polymorphs shows some differences.

So, for the YIS-1 and YIS-3 polymorphs a series of high-dispersive Ir 5$d$ - like bands intersects the Fermi level, whereas for YIS-2 simultaneously with a set of high-dispersive bands, the quasi-flat bands along R-X and A-M appear near E$_F$. These features yield a very distinct multi-sheet topology of the Fermi surface (FS) for each polymorph, see Fig. 4. Namely, the Fermi surface of the CaBe$_2$Ge$_2$-like YIS-2 consists of a set of quasi-two-dimensional hole and electronic pockets, which seem similar to the FSs for related 122-like Fe-based materials [2-12]. On the contrary, the Fermi surfaces of both YIS-1 and YIS-3 polymorphs differ essentially from those of the Fe-based 122 materials and are of a characteristic multi-sheet three-dimensional type like the ThCr$_2$Si$_2$-like iron-free $AM_2$As$_2$ phases [2-12].

The total, atomic, and orbital decomposed partial DOSs at the Fermi level, N(E$_F$), are shown in Table 4. It is seen that for all YIr$_2$Si$_2$ polymorphs the main contribution to N(E$_F$) comes from the Ir 5$d$ states, with some additions of the Y 4$d$- and Si 3$p$ states. For the examined polymorphs, the values of N(E$_F$) increase in the sequence: YIS-1 > YIS-2 > YIS-3. The obtained data also allowed us to estimate the Sommerfeld constants ($\gamma$) and the Pauli paramagnetic susceptibility ($\chi$) for YIr$_2$Si$_2$ polymorphs under the assumption of the free electron model as $\gamma = (\pi^2/3)N(E_F)k_B^2$ and $\chi = \mu_B^2 N(E_F)$, Table 4.

Note also that for the CaBe$_2$Ge$_2$-type polymorph (YIS-2) the contributions to N(E$_F$) from the states of various blocks ([Ir$_2$Si$_2$] *versus* [Si$_2$Ir$_2$]) differ appreciably. So, the value of N$^{Ird}$(E$_F$) = 0.250 states/eV·atom for Ir$^1$ atoms (placed inside [Ir$_2$Si$_2$] blocks) is much smaller than N$^{Ird}$(E$_F$) = 0.391 states/eV·atom for Ir$^2$ atoms located on the outer sides of [Si$_2$Ir$_2$] blocks. Therefore, the conduction in YIS-2 is expected to be anisotropic, *i.e.* happening mainly in the [Si$_2$Ir$_2$] blocks.

*3.4. Inter-atomic bonding*

A conventional picture of inter-atomic interactions in 122-like $AM_2$As$_2$ phases assumes [2-12] strong $M$-As bonding of a mixed ionic-covalent type inside [$M_2$As$_2$] blocks, some covalent As-As interactions between the adjacent [$M_2$As$_2$]/[$M_2$As$_2$] blocks together with ionic bonding between [$M_2$As$_2$] blocks and atomic $A$ sheets.

In our case, the overall character of Ir-Si bonding in the YIr$_2$Si$_2$ polymorphs **inside** [Ir$_2$Si$_2$] (or [Si$_2$Ir$_2$]) blocks may be understood from site-projected DOS calculations. As is shown in Fig. 3, the Ir 5$d$ and Si 3$p$ states are strongly hybridized. On the other hand, a completely different bonding in the YIr$_2$Si$_2$ polymorphs arises **between** the adjacent blocks, and this situation is clearly visible in Fig. 5. So, while for YIS-2 strong covalent Ir-Si bonding takes place, for the other polymorphs, directed unipolar Si-Si (YIS-1) or Ir-Ir bonds (YIS-3) appear. Besides, for the most stable YIS-1 and YIS-2 polymorphs, direct Y-Ir and Y-Si bonds are visible (Fig. 5), whereas for YIS-3 the Y-Si interactions are very weak. This factor may be responsible for the reduction in the stability of this polymorph.

Thus, summarizing the above results, the inter-atomic bonding for the $YIr_2Si_2$ polymorph phases can be classified as an anisotropic mixture of covalent, metallic, and ionic contributions, where inside the $[Ir_2Si_2]$ (or $[Si_2Ir_2]$) blocks, mixed covalent-metallic-ionic Ir-Si and Ir-Ir bonds take place (owing to hybridization of Ir $5d$ – Si $3p$ states, delocalized near-Fermi Ir $5d$ states, and Ir → Si charge transfer, respectively), whereas between the adjacent $[Ir_2Si_2]$ (or $[Si_2Ir_2]$) blocks and Y atomic sheets, for various polymorphs, Si-Si and Ir-Y (for YIS-1), Si-Ir and Si-Y (for YIS-2) as well as mainly Ir-Ir bonds (for YIS-3) emerge.

## 4. Conclusions

In this work, by means of the two complementary first-principles approaches (FLAPW and VASP codes), the structural, elastic, and electronic properties of three polymorphs of the superconducting disilicide $YIr_2Si_2$, which differ in the atomic configurations of $[Ir_2Si_2]$ (or $[Si_2Ir_2]$) blocks, were examined.

Our studies showed that both within the FLAPW and VASP calculations, the LT $ThCr_2Si_2$-type polymorph is the most stable, whereas the HT phase ($CaBe_2Ge_2$-like) is by about 0.1 eV/f.u. less stable. These polymorphs belong also to mechanically stable materials. On the contrary, the hypothetical "intermediate" $YIr_2Si_2$ polymorph becomes the most energetically unstable system and is also mechanically unstable.

We found also that the $ThCr_2Si_2$- and $CaBe_2Ge_2$-type polymorphs being relatively hard materials with low compressibility will behave as ductile systems. Among them the $ThCr_2Si_2$-type polymorph will show enhanced elastic anisotropy.

In the vicinity of the Fermi energy, the topology of the electronic bands and Fermi surface for various polymorphs is quite different. Besides, the $CaBe_2Ge_2$-type polymorph is expected to be anisotropic, *i.e.* happening mainly in the $[Si_2Ir_2]$ blocks.

The inter-atomic bonding for the $YIr_2Si_2$ polymorph phases can be described as an anisotropic mixture of covalent, metallic, and ionic contributions, where inside the $[Ir_2Si_2]$ (or $[Si_2Ir_2]$) blocks, Ir-Si and Ir-Ir bonds take place, whereas between the adjacent $[Ir_2Si_2]$ (or $[Si_2Ir_2]$) blocks and Y atomic sheets, Si-Si and Ir-Y, Si-Ir and Si-Y, or mainly Ir-Ir bonds emerge for various polymorphs.

## Acknowledgements

Financial support from the RFBR (Grants 09-03-00946 and 10-03-96008-Ural) is gratefully acknowledged.

# References

[1] Y. Kamihara, T. Watanabe, M. Hirano, H. Hosono, J. Am. Chem. Soc. 130 (2008) 3296.
[2] M.V. Sadovskii, Physics - Uspekhi 51 (2008) 1201.
[3] A.L. Ivanovskii, Physics - Uspekhi 51 (2008), 1229.
[4] Z.A. Ren, Z.X. Zhao, Adv. Mater. 21 (2009) 4584.
[5] J. A. Wilson, J. Phys.: Condens. Matter 22 (2010) 203201.
[6] A. L. Ivanovskii, Russ. Chem. Rev. 79 (2010) 1.
[7] D.C. Johnson, Adv. Phys. 59 (2010) 803.
[8] M. Rotter, M. Tegel, D. Johrendt, Phys. Rev. Lett. 101 (2008) 107006.
[9] M. Rotter, M. Tegel, D. Johrendt, I. Schellenberg, W. Hermes, R. Pöttgen, Phys. Rev. B 78 (2008) 020503.
[10] D. Kasinathan, A. Ormeci, K. Koch, U. Burkhardt, W. Schnelle, A. Leithe-Jasper, H. Rosner, New J. Phys. 11 (2009) 025023.
[10] C. Krellner, N. Caroca-Canales, A. Jesche, H. Rosner, A. Ormeci, C. Geibel, Phys. Rev. B 78 (2008) 100504.
[11] H. Hiramatsu, T. Kamiya, M. Hirano, H. Hosono, Physica C 469 (2009) 657.
[12] P. C. Canfield, S. L. Bud'ko, Ann. Rev. Cond. Matter Phys. 1 (2010) 27.
[13] J. Guo, S. Jin, G.Wang, S. Wang, K. Zhu, T. Zhou, M. He, X. Chen, Phys. Rev. B 82 (2010) 180520.
[14] A. Krzton-Maziopa, Z. Shermadini, E. Pomjakushina, V. Pomjakushin, M. Bendele, A. Amato, R. Khasanov, H. Luetkens, K. Conder, J. Phys.: Condens. Matter 23 (2011) 052203.
[15] I.R. Shein, A.L. Ivanovskii, Phys. Lett. A 375 (2011) 1028.
[16] T. Mine, H. Yanagi, T. Kamiya, Y. Kamihara, M. Hirano, H. Hosono, Solid State Commun. 147 (2008) 111.
[17] F. Ronning, N. Kurita, E.D. Bauer, B. L. Scott, T. Park, T. Klimczuk, R. Movshovich, J. D. Thompson, J. Phys.: Cond. Matter 20 (2008) 342203.
[18] E. D. Bauer, F. Ronning, B. L. Scott, J. D. Thompson, Phys. Rev. B 78 (2008) 172504.
[19] Y. Singh, Y. Lee, S. Nandi, A. Kreyssig, A. Ellern, S. Das, R. Nath, B. N. Harmon, A. I. Goldman, D. C. Johnston, Phys. Rev. B 78 (2008) 104512.
[20] I. R. Shein, A. L. Ivanovskii, Solid State Commun. 149 (2009) 1860.
[21] I. R. Shein, A. L. Ivanovskii, Phys. Rev. B 79 (2009) 054510.
[22] A. Imre, A. Hellmann, G. Wenski, J. Graf, D. Johrendt, A. Mewis, Z. Anorg. Allg. Chem. 633 (2007) 2037.
[23] K. Kudo, Y. Nishikubo, M. Nohara, Journal Phys. Soc. Japan, 79 (2010) 123710.
[24] W.K. Hoffmann, W. Jeitschko, J. Less Common Met. 138 (1988) 313.
[25] F. Wastin, J. Rebizant, J.C. Spirlet, C. Sari, C.T. Walker, J. Fuger, J. Allows Comp. 196 (1993) 87.

[26] R. Pottgen, J.H. Albering, D. Kaczorowski, W. Jeitschko, J. Allows Comp., 196 (1993) 111.
[27] S.K. Dhar, P. Paulose, R. Kulkarni, P. Manfrinetti, M. Pani, N. Parodi, Solid State Commun, 149 (2009) 68.
[28] R.N. Shelton, H.F. Braun, E. Musick, Solid State Commun.52 (1984) 797.
[29] S. Das, K. McFadden, Y. Singh, R. Nath, A. Ellern, D.C. Johnston, Phys. Rev. B 81 (2010) 054425.
[30] I. Higashi, P. Lejay, B. Chevalier, J. Etourneau, P. Hagenmuller, Rev. Chim. Miner., 21 (1984) 239.
[31] M. Hirjak, P. Lejay, B. Chevalier, J. Etourneau, P. Hagenmuller, J. Less Common Met. 105 (1985) 139.
[32] P. Blaha, K. Schwarz, G. K. H. Madsen, D. Kvasnicka, and J. Luitz, WIEN2k, *An Augmented Plane Wave Plus Local Orbitals Program for Calculating Crystal Properties,* (Vienna University of Technology, Vienna, 2001)
[33] J. P. Perdew, S. Burke, M. Ernzerhof, Phys. Rev. Lett. 77 (1996) 3865.
[34] P.E. Blőchl, O. Jepsen, O.K. Andersen, Phys. Rev. B 49 (1994) 16223.
[35] G. Kresse, D. Joubert, Phys. Rev. B 59 (1999) 1758.
[36] G. Kresse, J. Furthmuller, Phys. Rev. B 54 (1996) 11169.
[37] J.F. Nye, *Physical Properties of Crystals*, Oxford University Press, Oxford, 1985.
[38] Z. Wu, E. Zhao, H. Xiang, X. Hao, X. Liu, J. Meng, Phys. Rev. B 76 (2007) 054115.
[39] G. Grimvall, *Thermophysical Properties of Materials*, North-Holland, Amsterdam, 1986.
[40] I. R. Shein, A. L. Ivanovskii, Scripta Mater. 59 (2008) 1099.
[41] M. Mito, M. J. Pitcher, W. Crichton, G. Garbarino, P. J. Baker, S. J. Blundell, P. Adamson, D. R. Parker, S. J. Clarke, J. Am. Chem. Soc. 131 (2009) 2986.
[42] I. R. Shein, A. L. Ivanovskii, Tech. Phys. Lett. 35 (2009) 961.
[43] I. R. Shein, A. L. Ivanovskii, Physica C 469 (2009) 15 (2009).
[44] J. Haines, J.M. Leger, and G. Bocquillon, Annu. Rev. Mater. Res. 31 (2001) 1.
[45] S.F. Pugh, Philos. Mag. 45 (1953) 823.
[46] P. Ravindran, L. Fast, P.A. Korzhavyi, B. Johnnsson, J.Wills, O. Eriksson, J. Appl. Phys. 84 (1998) 4891.
[47] S. I. Ranganathan, M. Ostoja-Starzewshi, Phys. Rev. Lett. 101 (2008) 055504.

**Table 1.** The optimized lattice parameters ($a$ and $c$, in Å), internal coordinates ($z_{Y,Ir,Si}$), some inter-atomic distances ($d$, in Å), and total-energy differences ($\Delta E$, eV/form.unit) for the examined YIr$_2$Si$_2$ polymorphs.

| phase/parameter [1] | YIS-1 | YIS-2 | YIS-3 |
|---|---|---|---|
| $a$ | 4.0862/4.0817 (4.047) [2] | 4.1343/4.1365 (4.081) [2] | 4.1543/4.1739 |
| $c$ | 10.0008/9.9848 (9.914) | 9.9103/9.8487 (9.754) | 10.1334/9.9870 |
| $c/a$ | 2.4475/2.4462 (2.449) | 2.3971/2.3809 (2.390) | 2.4392/2.3927 |
| $z_{Y,Ir,Si}$ | $z_{Si}$=0.3785/0.3712 | $z_{Y}$=0.2543/0.2506 $z_{Ir}$=0.3709/0.3678 $z_{Si}$=0.1307/0.1272 | $z_{Ir}$=0.3715/0.3694 |
| $d^1$ | 2.41/2.41 (Ir-Si in [Ir$_2$Si$_2$]) 3.87/3.87 (Si-Si in [Ir$_2$Si$_2$]) | 2.44/2.44 (Ir-Si in [Ir$_2$Si$_2$]) 2.43/2.42 (Ir-Si in [Si$_2$Ir$_2$]) | 2.41/2.40 (Ir-Si in [Si$_2$Ir$_2$]) 3.82/3.80 (Ir-Ir in [Si$_2$Ir$_2$]) |
| $d^2$ | 2.44/2.42 (Si-Si) | 2.38/2.37 (Ir-Si) | 2.61/2.61 (Ir-Ir) |
| $\Delta E$ | 0/0 | 0.12/0.11 | 1.04/0.96 |

[1] as obtained within FLAPW/VASP
[2] available experimental data (Ref. [28]) are given in parentheses.
$d^1$ are the inter-atomic distances inside [Ir$_2$Si$_2$] ([Si$_2$Ir$_2$]) blocks.
$d^2$ are the nearest Si-Si (Si-Ir, or Ir-Ir) distances between neighboring blocks: [Ir$_2$Si$_2$]/[Ir$_2$Si$_2$] for YIS-1 ([Ir$_2$Si$_2$]/[Si$_2$Ir$_2$] for YIS-2, or [Si$_2$Ir$_2$]/[Si$_2$Ir$_2$] for YIS-3).

**Table 2**. Calculated elastic constants ($C_{ij}$, in GPa) for the examined YIr$_2$Si$_2$ polymorphs.

| phase/parameter * | YIS-1 | YIS-2 | YIS-3 |
|---|---|---|---|
| $C_{11}$ | 313 | 259 | 207 |
| $C_{12}$ | 130 | 149 | 144 |
| $C_{13}$ | 98 | 139 | 155 |
| $C_{33}$ | 305 | 281 | 219 |
| $C_{44}$ | 99 | 66 | 41 |
| $C_{66}$ | 14 | 66 | < 0 |

* as obtained within VASP

**Table 3.** Calculated elastic parameters for two $YIr_2Si_2$ polymorphs: bulk moduli ($B$, in GPa), compressibility ($\beta$, in GPa$^{-1}$), shear moduli ($G$, in GPa), Pugh's indicator ($G/B$), Young's moduli ($Y$, in GPa), Poisson's ratio ($\nu$), and the so-called universal anisotropy index ($A^U$).

| Phase/parameter * | YIS-1 | YIS-2 |
|---|---|---|
| $B_V$ | 175.9 | 183.7 |
| $B_R$ | 175.4 | 183.6 |
| $B_{VRH}$ | 175.6 | 183.6 |
| $\beta$ | 0.005694 | 0.005446 |
| $G_V$ | 82.7 | 64.4 |
| $G_R$ | 45.3 | 68.2 |
| $G_{VRH}$ | 64.0 | 66.3 |
| $G/B$ | 0.365 | 0.361 |
| $Y$ | 171.2 | 177.6 |
| $\nu$ | 0.3375 | 0.3388 |
| $|A^U|$ | 0.84 | 0.05 |

\* as obtained within VASP

**Table 4**. Total (in states/eV·form.unit) and partial (in states/eV·atom) densities of states at the Fermi level, electronic heat capacity $\gamma$ (in mJ·K$^{-2}$·mol$^{-1}$), and molar Pauli paramagnetic susceptibility $\chi$ (in $10^{-4}$ emu·mol$^{-1}$) for the examined $YIr_2Si_2$ polymorphs, as obtained within FLAPW.

| Parameter | YIS-1 | YIS-2 | YIS-3 | parameter | YIS-1 | YIS-2 | YIS-3 |
|---|---|---|---|---|---|---|---|
| Total | 2.108 | 2.316 | 2.678 | $Ir^2\,5d$ | - | 0.391 | 0.417 |
| Y $5s$ | 0.003 | 0.002 | 0.002 | $Ir^2\,5d_z^2$ | - | 0.126 | 0.053 |
| Y $5p$ | 0.013 | 0.013 | 0.014 | $Ir^2\,5d_{xy}$ | - | 0.064 | 0.115 |
| Y $4d$ | 0.306 | 0.288 | 0.280 | $Ir^2\,5d_{x^2+y^2}$ | - | 0.053 | 0.068 |
| $Ir^1\,6s$ * | 0.006 | 0.005 | - | $Ir^2\,5d_{xz+yz}$ | - | 0.147 | 0.181 |
| $Ir^1\,6p$ | 0.024 | 0.031 | - | $Si^1\,3s$ | - | 0.005 | 0.011 |
| $Ir^1\,5d$ | 0.319 | 0.250 | - | $Si^1\,3p$ | - | 0.119 | 0.096 |
| $Ir^1\,5d_z^2$ | 0.045 | 0.014 | - | $Si^1\,3d$ | - | 0.018 | 0.016 |
| $Ir^1\,5d_{xy}$ | 0.023 | 0.026 | - | $Si^2\,3s$ | 0.006 | 0.013 | - |
| $Ir^1\,5d_{x^2+y^2}$ | 0.069 | 0.043 | - | $Si^2\,3p$ | 0.062 | 0.058 | - |
| $Ir^1\,5d_{xz+yz}$ | 0.182 | 0.167 | - | $Si^2\,3d$ | 0.028 | 0.021 | - |
| $Ir^2\,6s$ | - | 0.005 | 0.013 | $\chi$ | 0.679 | 0.746 | 0.862 |
| $Ir^2\,6p$ | - | 0.090 | 0.073 | $\gamma$ | 4.969 | 5.459 | 6.312 |

\* $(Ir,Si)^{1,2}$ – the non-equivalent atoms, see text.

**Figures**

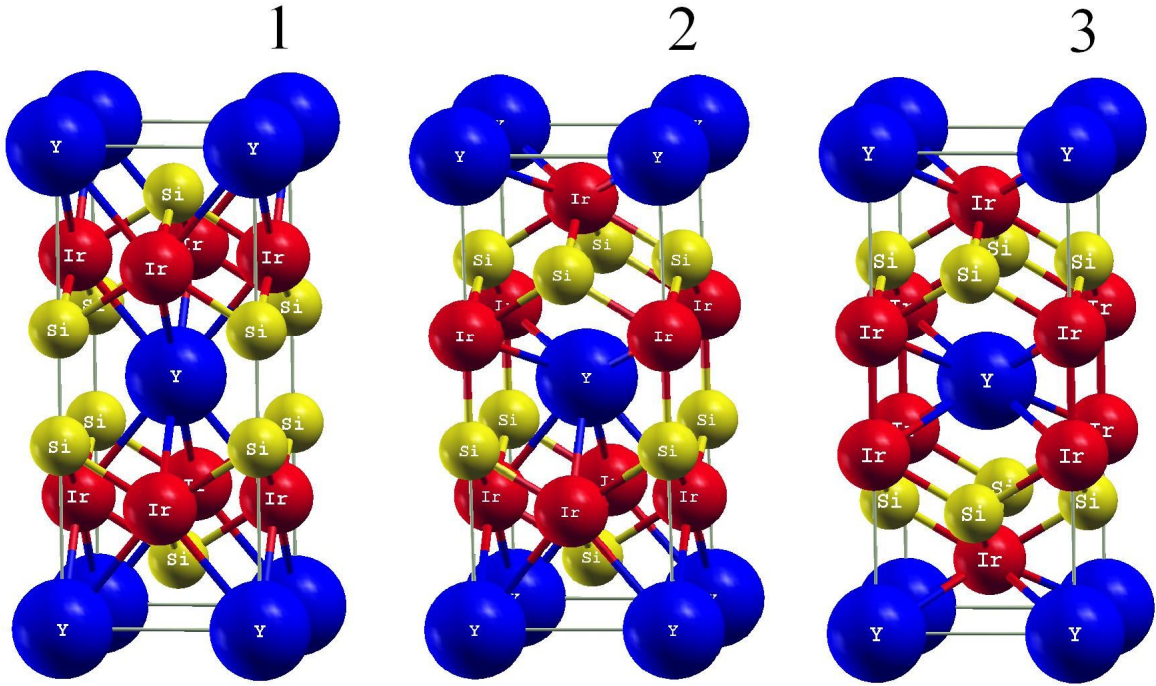

**Fig. 1.** (*Color online*) Crystal structures of YIr$_2$Si$_2$ polymorphs: (1) YIS-1, (2) YIS-2, and (3) YIS-3, *see text*.

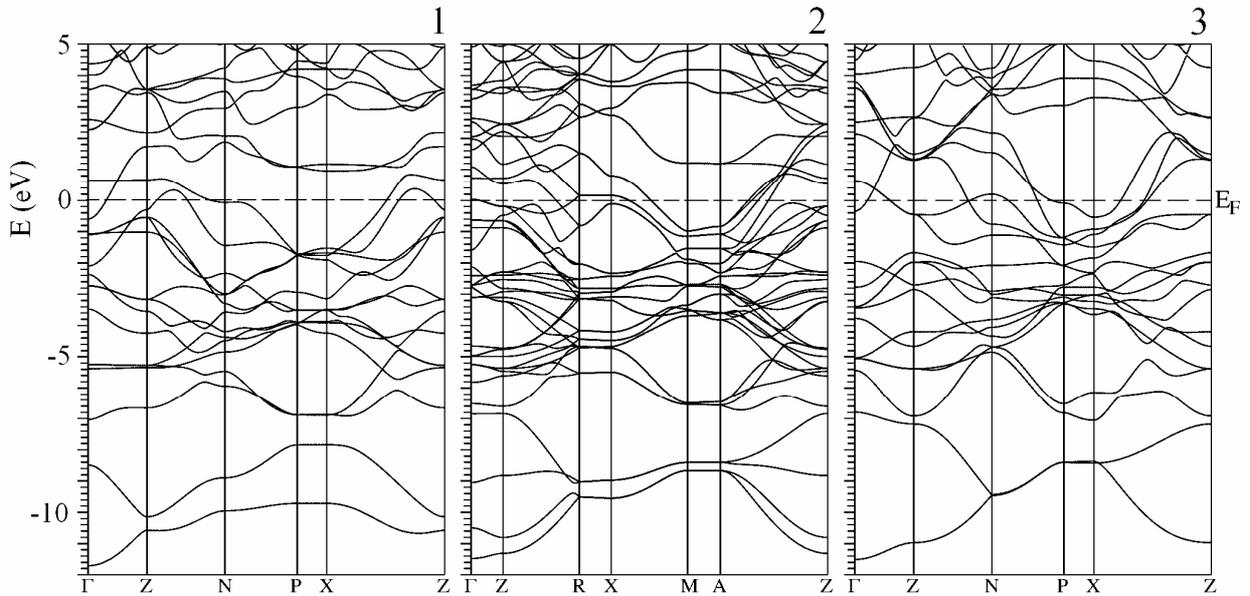

**Fig. 2**. Electronic band structures of YIr$_2$Si$_2$ polymorphs: (1) YIS-1, (2) YIS-2, and (3) YIS-3, *see Fig. 1*.
.

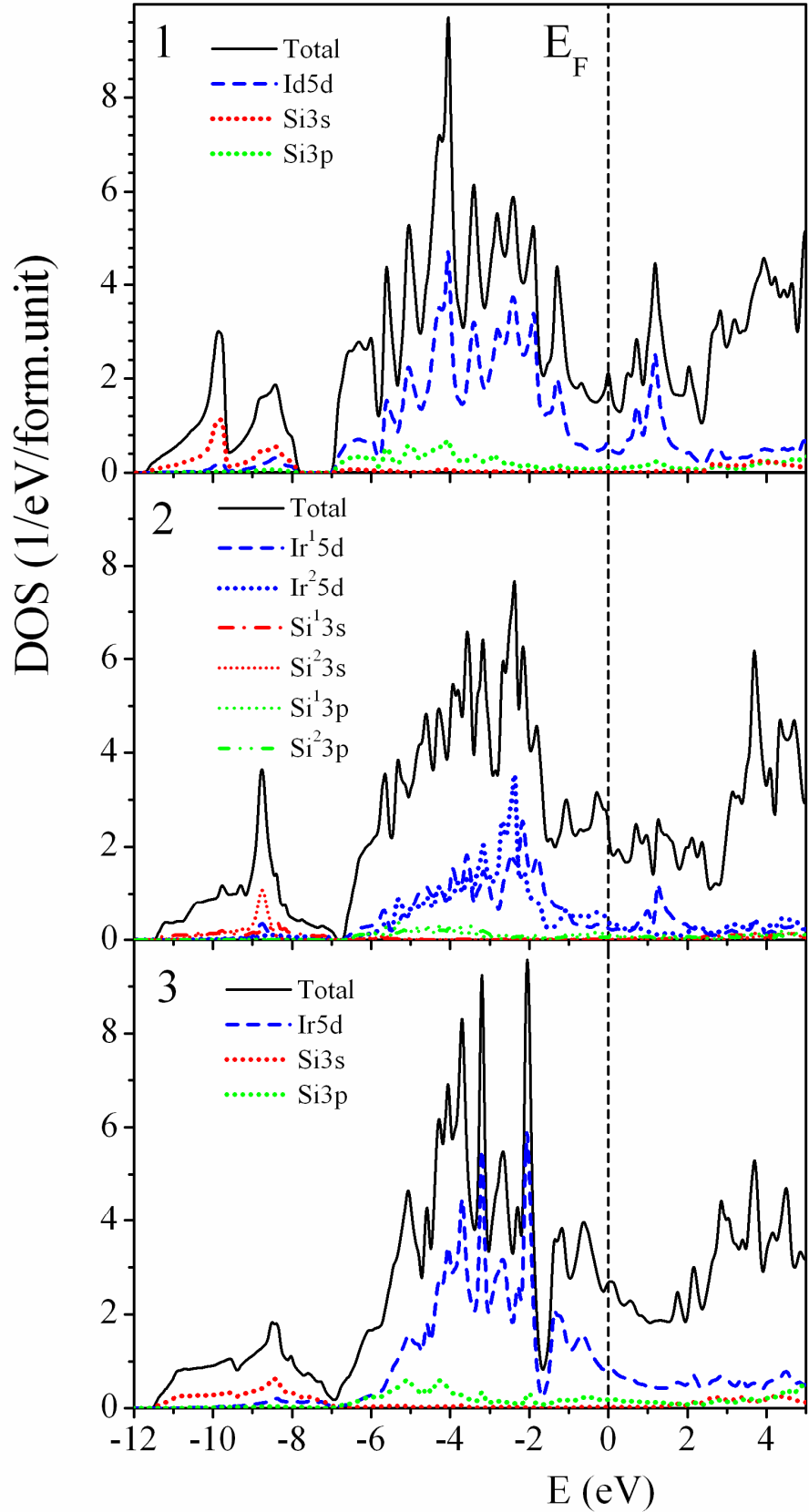

**Fig. 3.** (*Color online*) Total and partial densities of states of YIr$_2$Si$_2$ polymorphs: (1) YIS-1, (2) YIS-2, and (3) YIS-3, *see Fig. 1*.

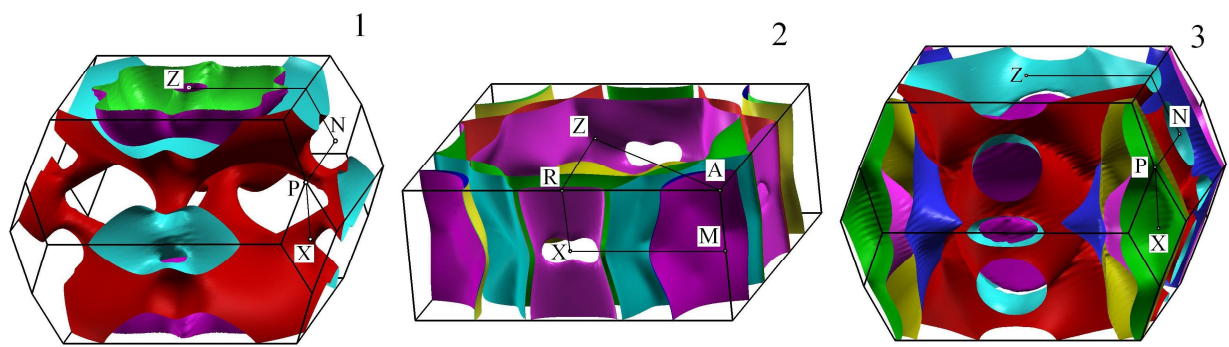

**Fig. 4.** (*Color online*) The Fermi surfaces of YIr$_2$Si$_2$ polymorphs: (1) YIS-1, (2) YIS-2, and (3) YIS-3, *see Fig. 1*

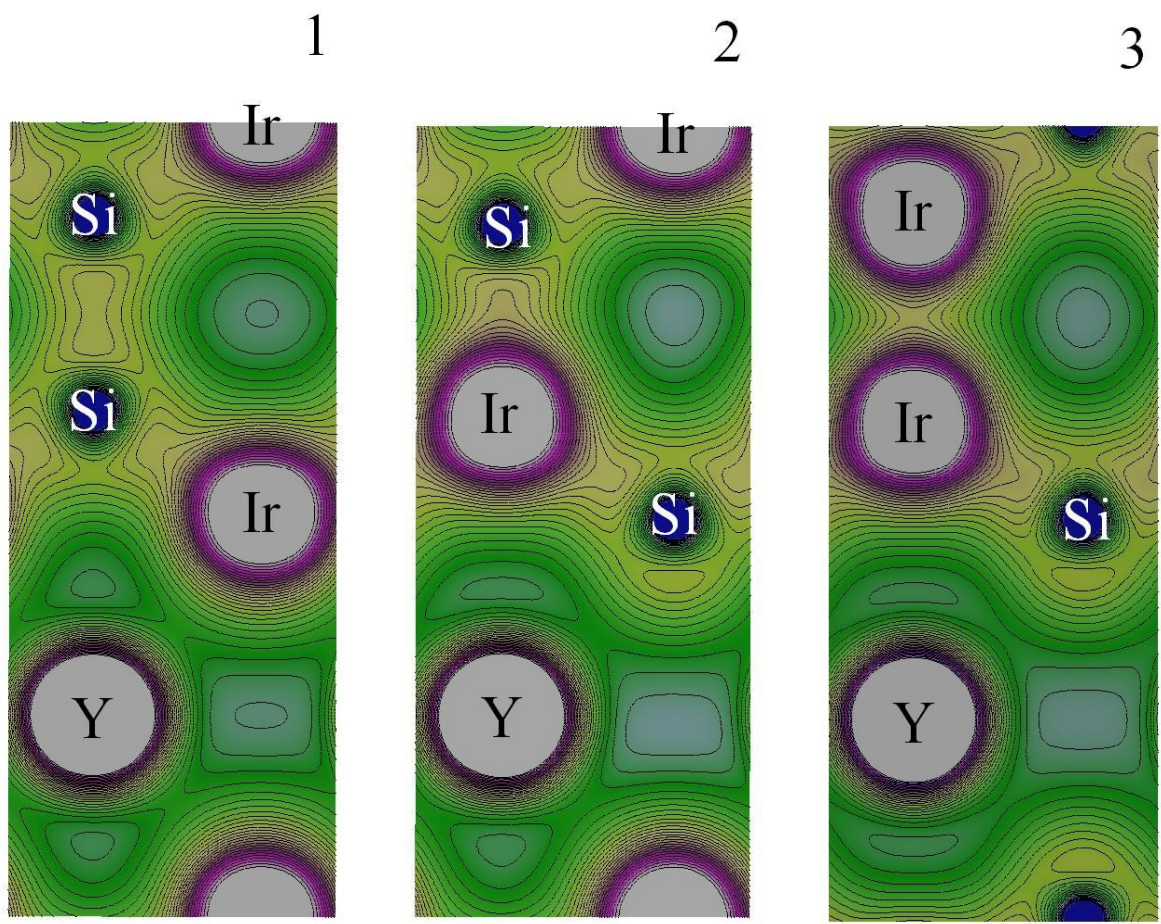

**Fig. 5**. (*Color online*) Charge density maps of YIr$_2$Si$_2$ polymorphs illustrating the formation of directional "inter-blocks" covalent bonds: (1) Si-Si for YIS-1, (2) Si-Ir for YIS-2, and (3) Ir-Ir for YIS-3.